\documentclass[epjCONF]{svjour}
\usepackage{graphics}
\usepackage{epsfig}
\usepackage[varg]{txfonts} 
\usepackage[latin1]{inputenc}
      \usepackage{bm}
\session-title{MESON2012 - 12th International Workshop on Meson Production, Properties and Interaction}
\begin{document}
\title{Open charm meson production at LHC}
\author{Rafal Maciula\inst{1}\fnmsep\thanks{\email{rafal.maciula@ifj.edu.pl}} \and Marta Luszczak \inst{2}\fnmsep\thanks{\email{luszczak@univ.rzeszow.pl}} \and Antoni Szczurek\inst{1,2}\fnmsep\thanks{\email{antoni.szczurek@ifj.edu.pl}}\fnmsep\thanks{This work was supported in part by the Polish MNiSW grants No. N N202 237040 and DEC-2011/01/B/ST2/04535.} }
\institute{Institute of Nuclear Physics PAN, PL-31-342 Cracow, Poland \and University of Rzesz\'ow, PL-35-959 Rzesz\'ow, Poland}
\abstract{
We discuss charm production at the LHC. The production of single $c \bar c$ pairs
is calculated in the $k_t$-factorization approach. We use Kimber-Martin-Ryskin
unintegrated gluon distributions in the proton. The hadronization is included with the help of
Peterson fragmentation functions. Transverse momentum and pseudorapidity distributions of charmed mesons are presented and compared to recent results of the ALICE,
LHCb and ATLAS collaborations. Furthermore we discuss production of two pairs of $c \bar c$ within a simple
formalism of double-parton scattering (DPS). Surprisingly large
cross sections, comparable to single-parton scattering (SPS), are predicted for LHC energies.
We discuss perspectives how to identify the double scattering
contribution. We predict much larger cross section for large rapidity distance
between charm quarks from different hard parton scatterings
compared to single scattering.
} 
\maketitle
%

\section{Transverse momentum spectra of open charm mesons at LHC}

Recently ALICE, LHCb and ATLAS collaborations have measured inclusive transverse momentum spectra
of open charm mesons in proton-proton collisions 
at $\sqrt{s}=7$ TeV \cite{ALICE,LHCb,ATLAS}. These measurements are very
interesting from the theoretical point of view because of the collision
energy never achieved before and unique rapidity acceptance of the
detectors. Especially, results from forward rapidity region $2 < y < 4$,
obtained by the LHCb as well as ATLAS data from wide pseudorapidity
range $|\eta| < 2.1$ can improve our understanding of pQCD production of heavy quarks.

The inclusive production of heavy quark/antiquark pairs can be calculated
in the framework of the $k_t$-factorization \cite{CCH91}.
In this approach transverse momenta of initial partons are included and
emission of gluons is encoded in the so-called unintegrated gluon,
in general parton, distributions (UGDFs).
In the leading-order approximation (LO) within the $k_t$-factorization approach
the differential cross section for the $Q \bar Q$
can be written as:
\begin{eqnarray}
\frac{d \sigma}{d y_1 d y_2 d^2 p_{1t} d^2 p_{2t}} =
\sum_{i,j} \; \int \frac{d^2 k_{1,t}}{\pi} \frac{d^2 k_{2,t}}{\pi}
\frac{1}{16 \pi^2 (x_1 x_2 s)^2} \; \overline{ | {\cal M}_{ij} |^2}\\
\nonumber
\times \;\; \delta^{2} \left( \vec{k}_{1,t} + \vec{k}_{2,t} 
                 - \vec{p}_{1,t} - \vec{p}_{2,t} \right) \;
{\cal F}_i(x_1,k_{1,t}^2) \; {\cal F}_j(x_2,k_{2,t}^2) \; , \nonumber \,\,
\end{eqnarray}
where ${\cal F}_i(x_1,k_{1,t}^2)$ and ${\cal F}_j(x_2,k_{2,t}^2)$
are the unintegrated gluon (parton) distribution functions. 

There are two types of the LO $2 \to 2$ subprocesses which contribute
to heavy quarks production, $gg \to Q \bar Q$ and $q \bar q \to Q \bar
Q$. The first mechanism dominates at large energies and the second one
near the threshold. Only $g g \to Q \bar Q$ mechanism is included here. We use off-shell matrix elements
corresponding to off-shell kinematics so hard amplitude depends on transverse momenta
(virtualities of initial gluons). In the case of charm production at
very high energies, especially at forward rapidities, rather small
$x$-values become relevant. 
Taken wide range of $x$ necessary for the calculation
we follow the Kimber-Martin-Ryskin (KMR) \cite{KMR01} prescription for unintegrated gluon distributions. More details of theoretical model can be found in Ref.~\cite{LMS09}.

The hadronization of heavy quarks is usually done
with the help of fragmentation functions. The inclusive distributions of
hadrons can be obtained through a convolution of inclusive distributions
of heavy quarks/antiquarks and Q $\to$ h fragmentation functions:
\begin{equation}
\frac{d \sigma(y_h,p_{t,h})}{d y_h d^2 p_{t,h}} \approx
\int_0^1 \frac{dz}{z^2} D_{Q \to h}(z)
\frac{d \sigma_{g g \to Q}^{A}(y_Q,p_{t,Q})}{d y_Q d^2 p_{t,Q}}
\Bigg\vert_{y_Q = y_h \atop p_{t,Q} = p_{t,h}/z}
 \; ,
\label{Q_to_h}
\end{equation}
where $p_{t,Q} = \frac{p_{t,h}}{z}$, where $z$ is the fraction of longitudinal momentum of heavy quark carried by meson.
We have made approximation assuming that $y_{Q}$  is
unchanged in the fragmentation process.

\begin{figure}[!h]
\begin{center}
\begin{minipage}{0.47\textwidth}
 \centerline{\includegraphics[width=1.0\textwidth]{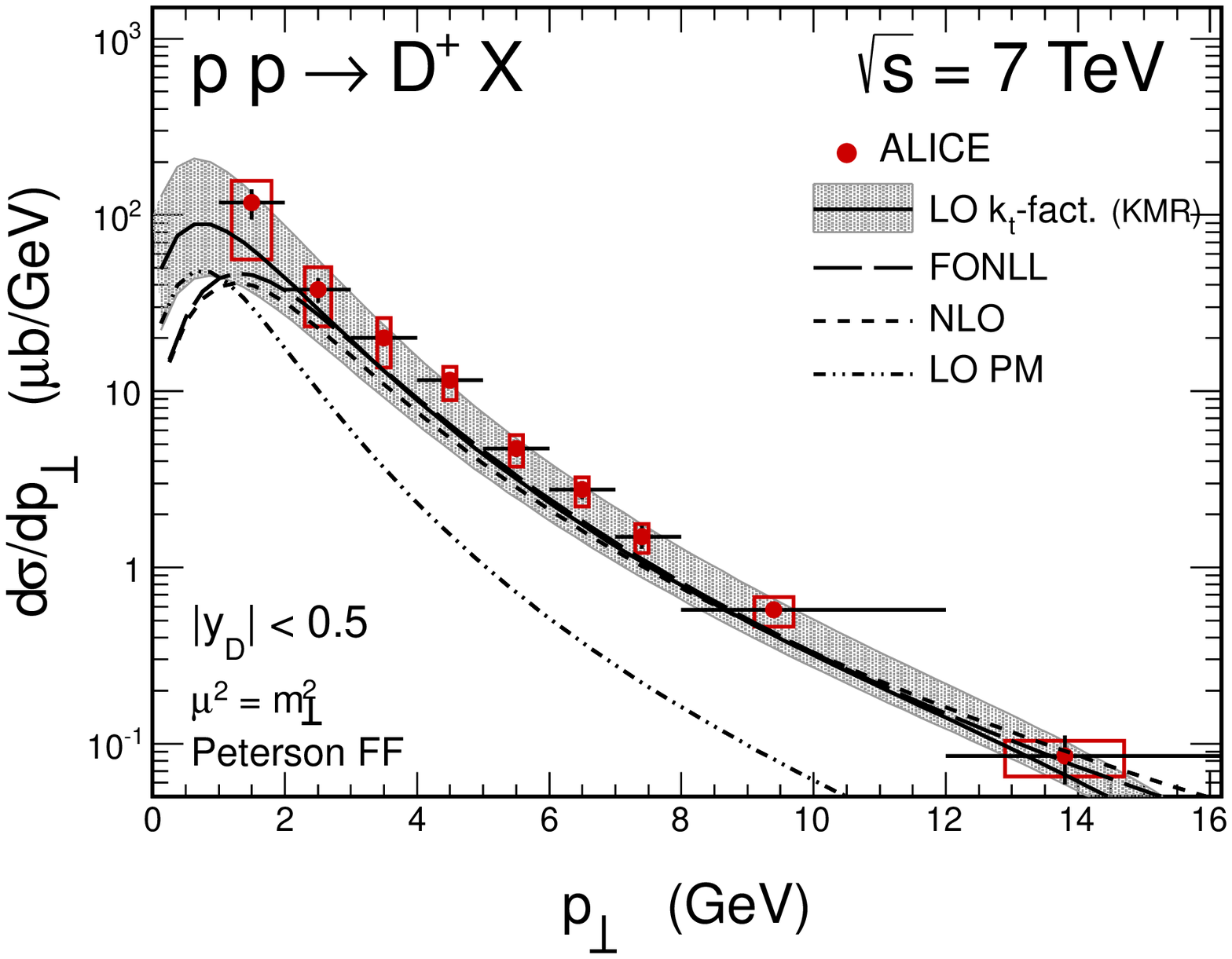}}
\end{minipage}
\hspace{0.5cm}
\begin{minipage}{0.47\textwidth}
 \centerline{\includegraphics[width=1.0\textwidth]{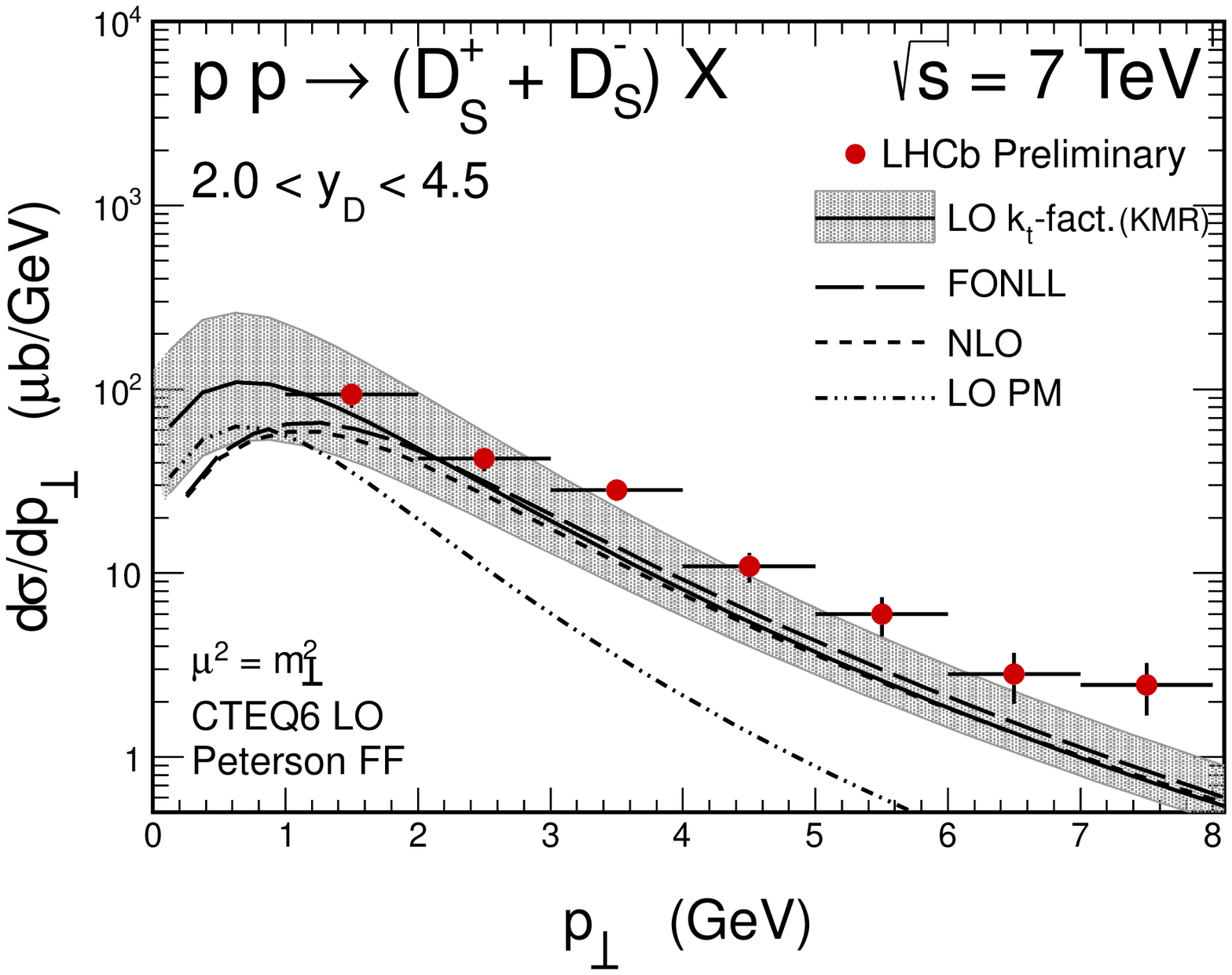}}
\end{minipage}
   \caption{
\small Transverse momentum distributions of $D^{\pm}$ and $D^{\pm}_{S}$ mesons together with the ALICE (left) and LHCb (right) data at $\sqrt{s} = 7$ TeV. Predictions of LO $k_t$-factorization together with uncertainties are compared with NLO parton model and FONLL calculations.}
 \label{fig:pt-alice-D-1}
\end{center}
\end{figure}

In Fig.~\ref{fig:pt-alice-D-1} we present our predictions for differential distributions in transverse momentum of open charm mesons together with
the ALICE (left panel) and LHCb (right panel) experimental data.
The uncertainties are obtained by changing charm quark mass $m_c = 1.5\pm 0.3$ GeV and by varying
renormalization and factorization scales $\mu^2=\zeta m_{t}^2$, where $\zeta \in (0.5;2)$. The gray shaded bands represent these both sources of uncertainties summed in quadrature. Using KMR model of UGDFs we get very good description of the experimental data, in both ALICE and LHCb cases. Here, we also compare central values of our LO $k_t$-factorization calculations (solid line) with NLO parton model (dashed line) and FONLL \cite{FONLL} predicitons (long-dashed line). All of these three models are consistent and give very similar results. The only difference is obtained at very small meson $p_{t}$'s (below $2$ GeV), where transverse momenta of initial gluons play a very improtant role.

\begin{figure}[!h]
\begin{minipage}{0.47\textwidth}
 \centerline{\includegraphics[width=1.0\textwidth]{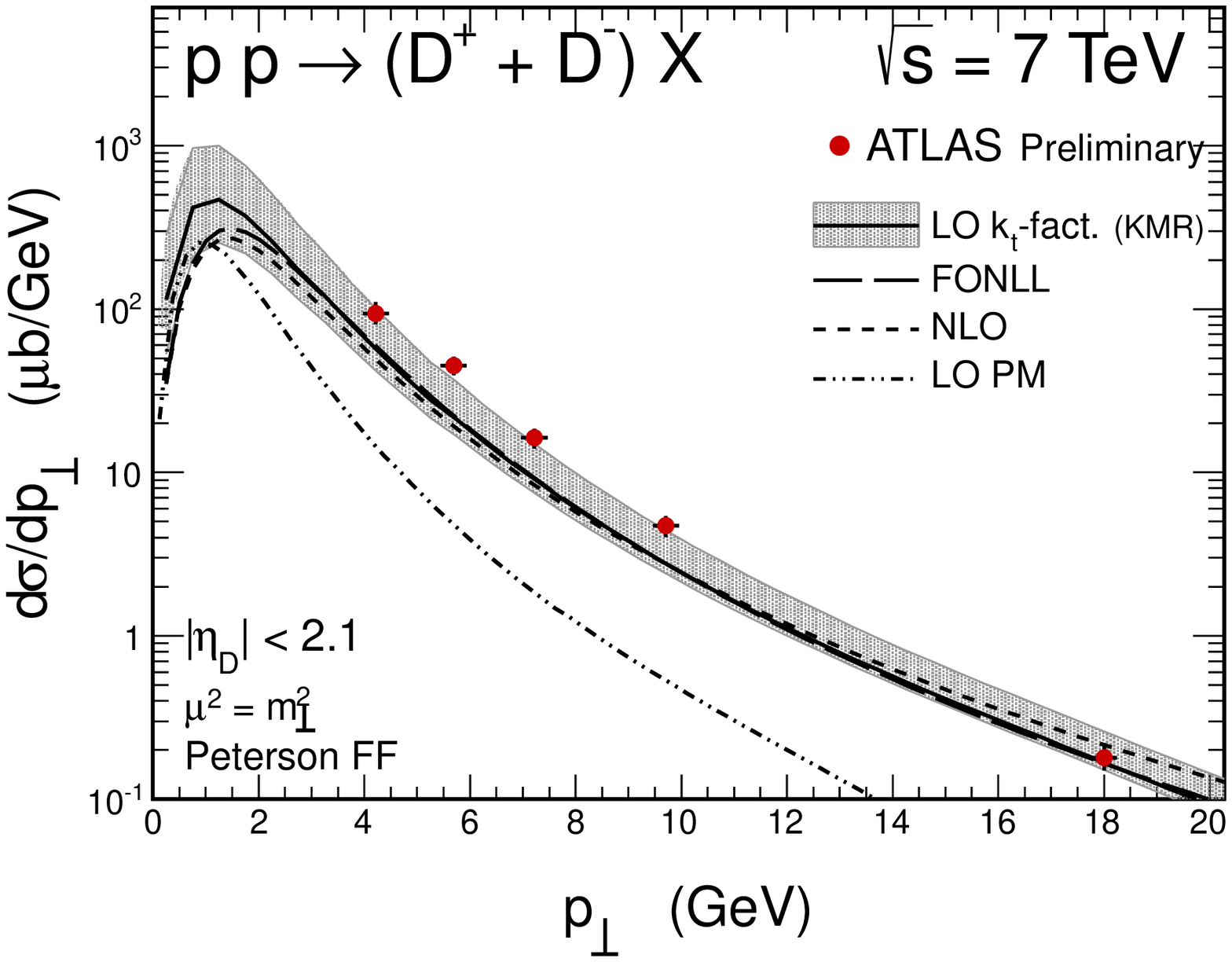}}
\end{minipage}
\hspace{0.5cm}
\begin{minipage}{0.47\textwidth}
 \centerline{\includegraphics[width=1.0\textwidth]{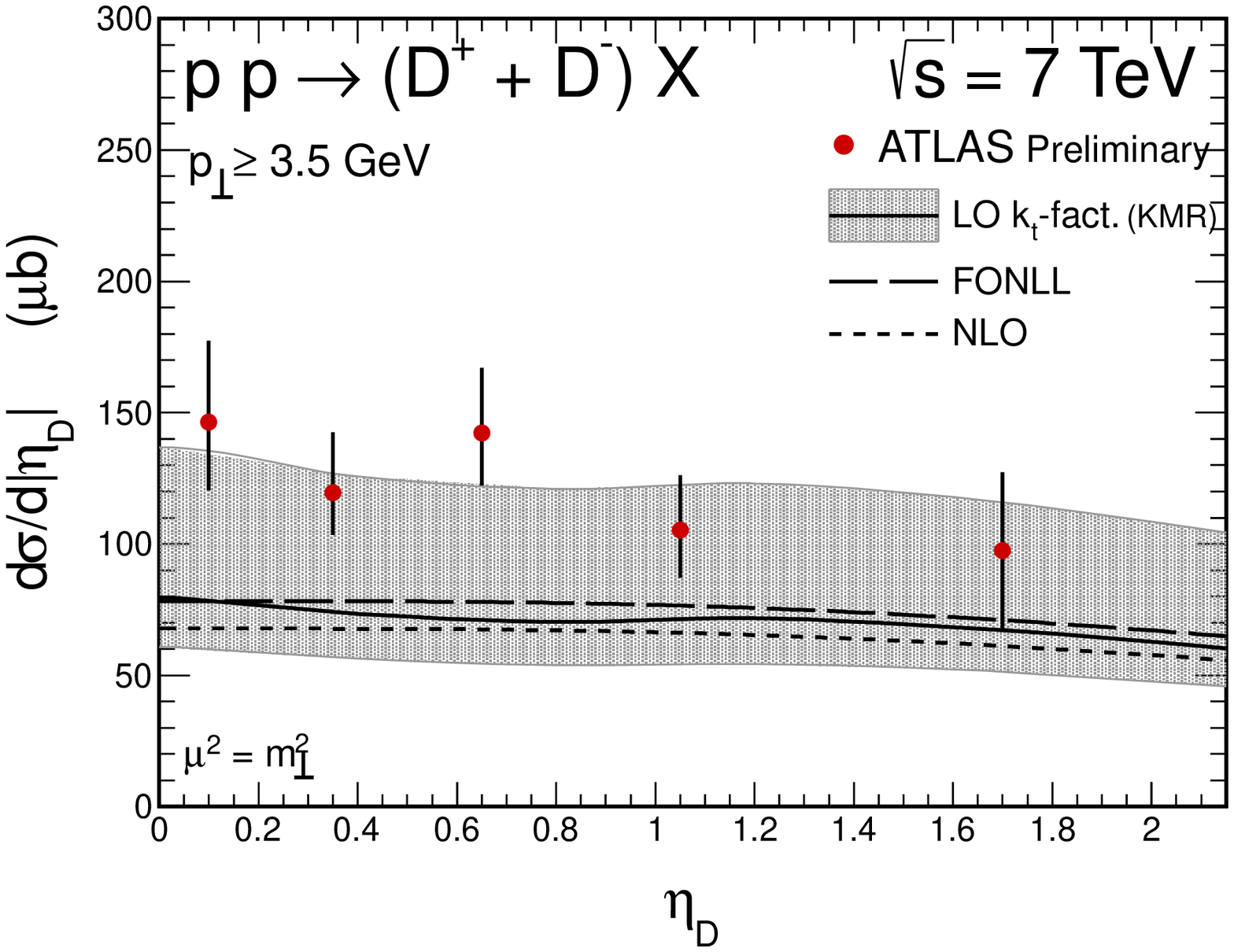}}
\end{minipage}
   \caption{
\small  Transverse momentum (left) and pseudorapidity (right)
distributions of $D^{\pm}$ mesons with the ATLAS experimental data at $\sqrt{s} = 7$ TeV. Predictions of LO $k_t$-factorization together with uncertainties are compared with NLO paton model and FONLL calculations.}
 \label{fig:pt-alice-D-2}
\end{figure}

In Fig.\ref{fig:pt-alice-D-2} we show transverse momentum (left panel) and pseudorapidity (right panel) spectra of $D^{\pm}$ mesons measured by ATLAS. The representation of theoretical results and uncertainties is the same as in Fig.~\ref{fig:pt-alice-D-1}. In contrast to the ALICE midrapidity measurements, here the experimental data points can be described only by the upper limit of our theoretical predictions. Therefore one can conclude, that covering wider range of (pseudo)rapidities (getting larger rapidity differences between produced quark and antiquark) the theoretical description of measured data becomes somewhat worse.

\section{Double charm production via Double Parton Scattering}

\begin{figure}[!h]
\begin{center}
\includegraphics[width=4cm]{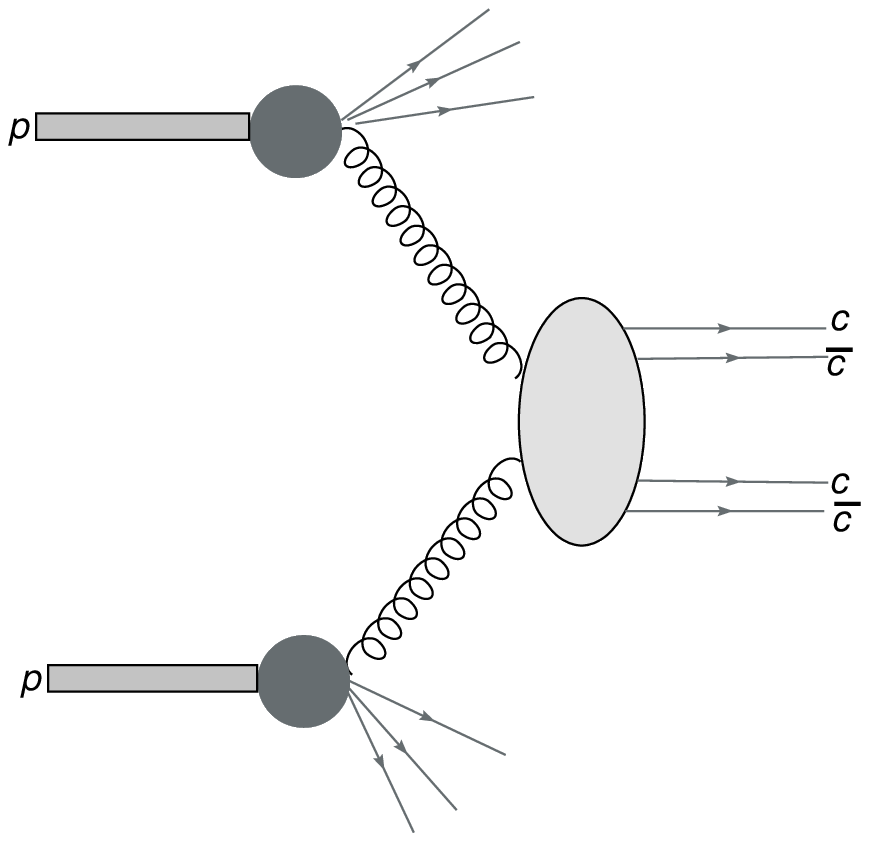}
\includegraphics[width=4cm]{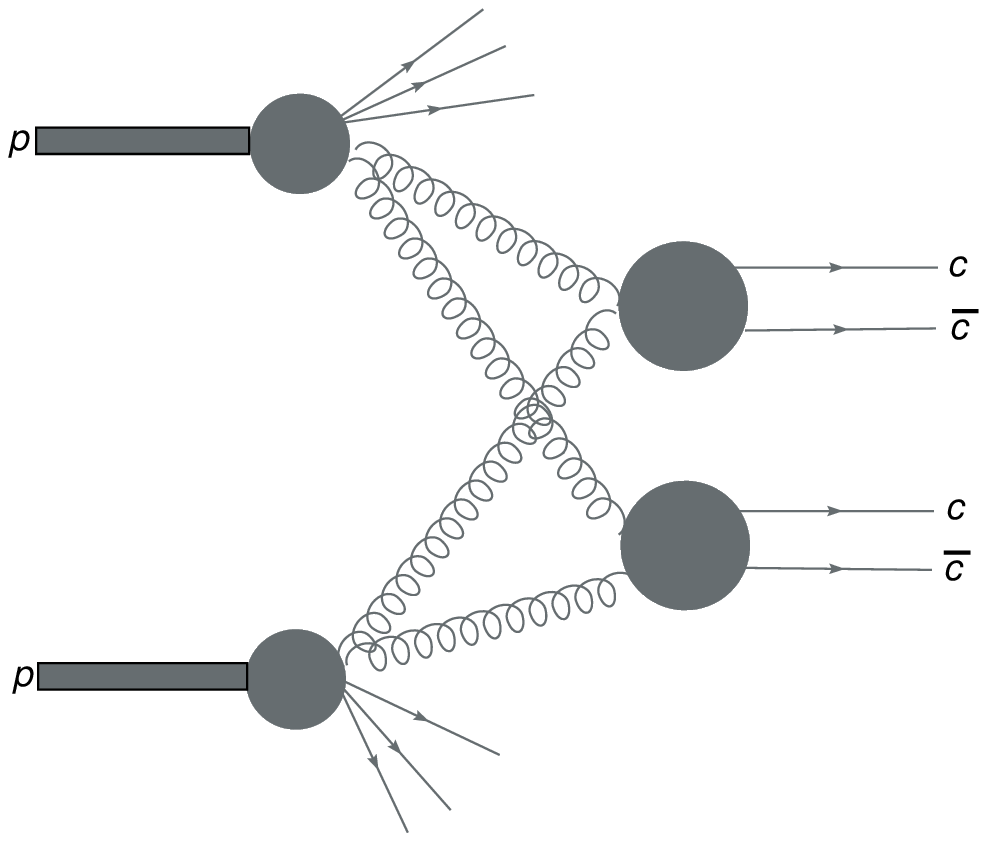}
\end{center}
   \caption{
\small SPS (left) and DPS (right) mechanisms of $(c \bar c) (c \bar c)$ 
production.  
}
 \label{fig:diagram}
\end{figure}

The mechanism of double-parton scattering (DPS) production of two pairs of
heavy quark and heavy antiquark is shown in Fig.~\ref{fig:diagram}
together with corresponding mechanism of single-scattering production.
The double-parton scattering has been recognized and discussed already 
in seventies and eighties. The activity stopped when it was realized that their contribution at center-of-mass energies available then was negligible. Nowadays, the theory of the double-parton
scattering is quickly developing (see e.g. \cite{S2003,KS2004,GS2010})
which is partly driven by new results from the LHC.

The double-parton scattering formalism in the simplest form assumes two
single-parton scatterings. Then in a simple probabilistic picture the
cross section for double-parton scattering can be written as:
\begin{equation}
\sigma^{DPS}(p p \to c \bar c c \bar c X) = \frac{1}{2 \sigma_{eff}}
\sigma^{SPS}(p p \to c \bar c X_1) \cdot \sigma^{SPS}(p p \to c \bar c X_2).
\label{basic_formula}
\end{equation}
This formula assumes that the two subprocesses are not correlated and do
not interfere.
At low energies one has to include parton momentum conservation
i.e. extra limitations: $x_1+x_3 <$ 1 and $x_2+x_4 <$ 1, where $x_1$ and $x_3$
are longitudinal momentum fractions of gluons emitted from one proton and $x_2$ and $x_4$
their counterparts for gluons emitted from the second proton. The "second"
emission must take into account that some momentum was used up in the "first" parton
collision. This effect is important at large quark or antiquark rapidities.
Experimental data \cite{Tevatron} provide an estimate of $\sigma_{eff}$
in the denominator of formula (\ref{basic_formula}). In our analysis we
take $\sigma_{eff}$ = 15 mb.

A more general formula for the cross section can be written formally 
in terms of double-parton distributions (dPDFs), e.g. $F_{gg}$, $F_{qq}$, etc. 
In the case of heavy quark production at high energies:
\begin{eqnarray}
d \sigma^{DPS} &=& \frac{1}{2 \sigma_{eff}}
F_{gg}(x_1,x_3,\mu_1^2,\mu_2^2) F_{gg}(x_2,x_4,\mu_1^2,\mu_2^2) \times
\nonumber \\
&&d \sigma_{gg \to c \bar c}(x_1,x_2,\mu_1^2)
d \sigma_{gg \to c \bar c}(x_3,x_4,\mu_2^2) \; dx_1 dx_2 dx_3 dx_4 \, .
\label{cs_via_doublePDFs}
\end{eqnarray}
It is physically motivated to write the dPDFs
rather in the impact parameter space 
$F_{gg}(x_1,x_2,b) = g(x_1) g(x_2) F(b)$, where $g$ are 
usual conventional parton distributions and $F(b)$ is an overlap of 
the matter distribution in the 
transverse plane where $b$ is a distance between both gluons
\cite{CT1999}. The effective cross section in 
(\ref{basic_formula}) is then $1/\sigma_{eff} = \int d^2b F^2(b)$ and 
in this approximation is energy independent.

\begin{figure}[!h]
\begin{minipage}{0.47\textwidth}
 \centerline{\includegraphics[width=1.0\textwidth]{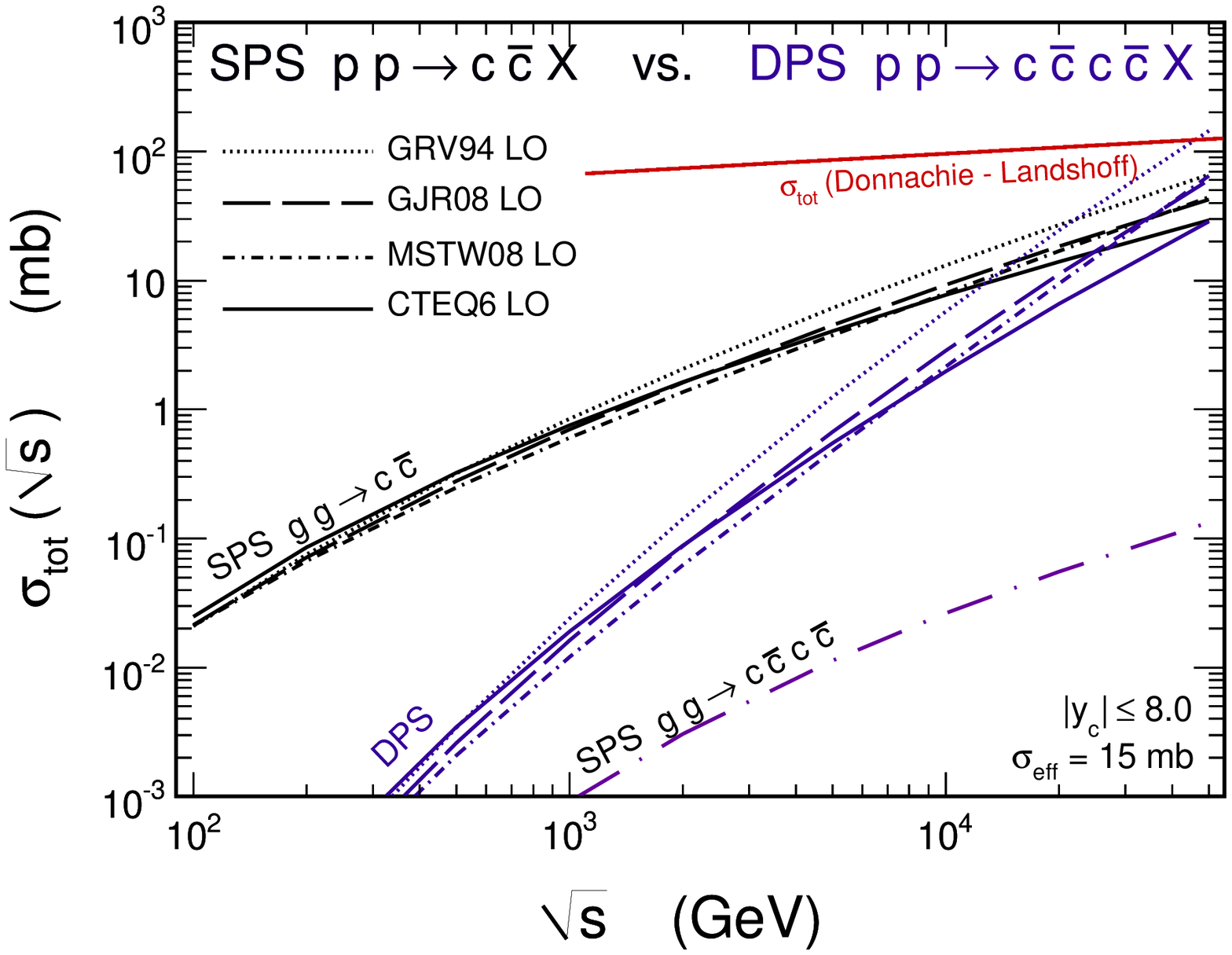}}
\end{minipage}
\hspace{0.5cm}
\begin{minipage}{0.47\textwidth}
 \centerline{\includegraphics[width=1.0\textwidth]{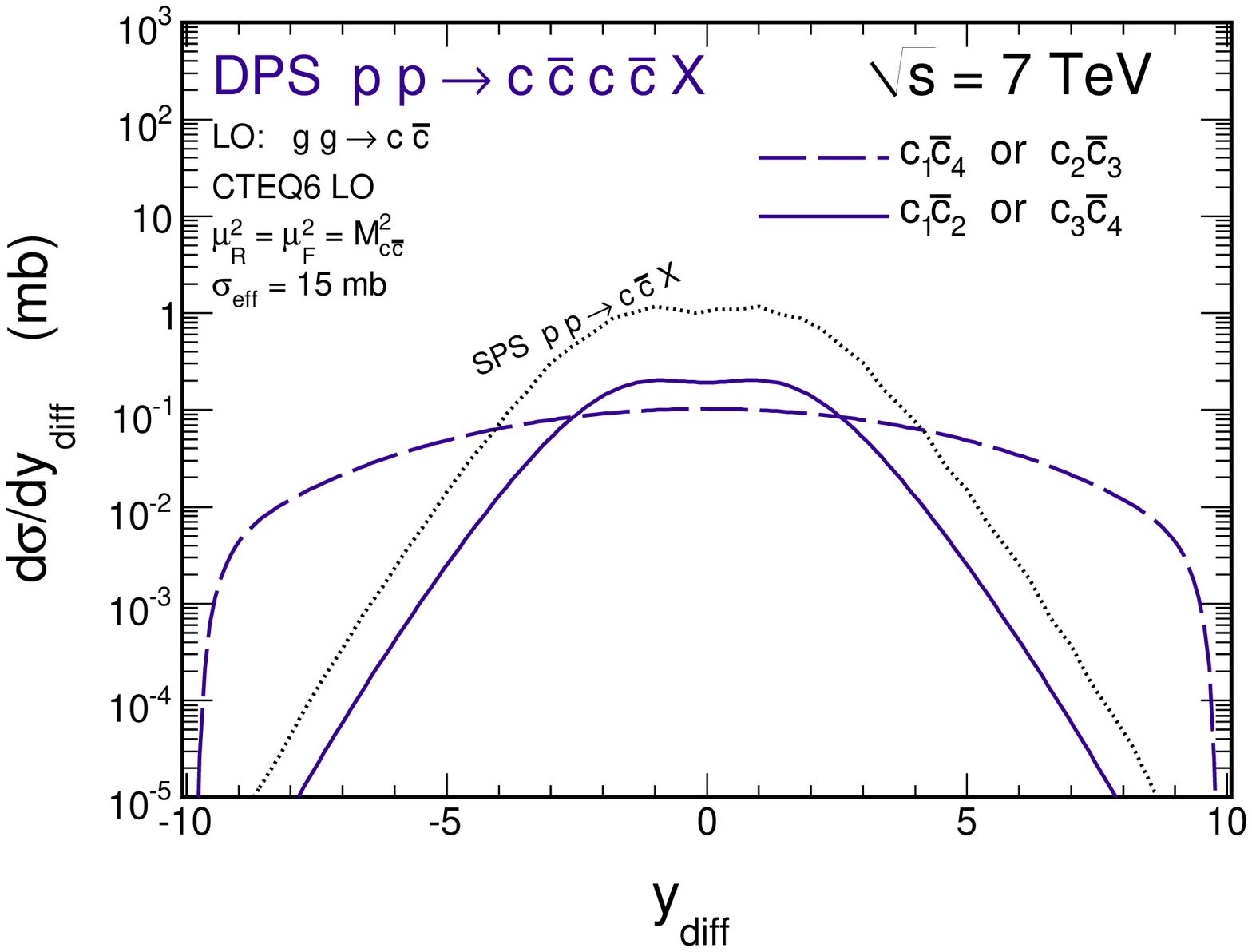}}
\end{minipage}
   \caption{
\small Total LO cross section for single
$c \bar c$ pair and SPS and DPS $c \bar c c \bar c$ production as a function of center-of-mass energy (left panel) and  
differential distribution in rapidity difference (right panel) between $c$ and $\bar{c}$ quarks  at $\sqrt{s}$ = 7 TeV.
Cross section for DPS should be multiplied in addition by a factor 2 in the case when all $c$ ($\bar c$) are counted.
We show in addition a parametrization of the total cross section in the left panel.
}
 \label{fig:single_vs_double_LO}
\end{figure}

In the left panel of Fig.~\ref{fig:single_vs_double_LO} we
compare cross sections for the single $c \bar c$ pair production as well
as for single-parton and double-parton scattering $c \bar c c \bar c$
production as a function of proton-proton center-of-mass energy. 
At low energies the single $c \bar c$ pair production
cross section is much larger. The cross section for SPS production
of $c \bar c c \bar c$ system \cite{SS2012} is more than two orders of magnitude smaller
than that for single $c \bar c$ production. For reference we show the
proton-proton total cross section as a function of energy as
parametrized in Ref.~\cite{DL92}. At low energy the $c \bar c$ or $ c \bar c c \bar c$ cross sections are much
smaller than the total cross section. At higher energies the contributions
approach the total cross section. This shows that inclusion of
unitarity effect and/or saturation of parton distributions may be necessary.
At LHC energies the cross section for both terms becomes comparable.
This is a new situation when the DPS gives a huge contribution to inclusive
charm production. 

In the right panel of Fig.~\ref{fig:single_vs_double_LO} we present
distribution in the difference of $c$ and $\bar c$ rapidities $y_{diff} = y_c - y_{\bar c}$.
We show both terms: when $c \bar c$ are emitted in the same parton scattering 
($c_1\bar c_2$ or $c_3\bar c_4$) and when they are emitted from different 
parton scatterings ($c_1\bar c_4$ or $c_2\bar c_3$). In the latter case
we observe a long tail for large rapidity difference as well as at large
invariant masses of $c \bar c$.

In particular, $c c$ (or $\bar c \bar c$) should be predominantly
produced from two different parton scatterings which opens a possibility 
to study the double scattering processes.
A good signature of the $c \bar c c \bar c$ final state is a production
of two mesons, both containing $c$ quark or two mesons
both containing $\bar c$ antiquark ($D^0 D^0$ or/and ${\bar D}^0 {\bar D}^0$) in one physical event.
A more detailed discussion of the DPS charm production can be found in
our original paper Ref.~\cite{LMS2012}.

In the present approach we have calculated cross section in a simple
collinear leading-order approach. A better approximation would be to include
multiple gluon emissions. This can be done e.g. in soft gluon resummation
or in $k_t$-factorization approach. This will be
discussed in detail elsewhere \cite{MS2012}.

\end{document}